\documentclass[twocolumn,prl,showpacs]{revtex4}
\usepackage{graphicx}
\usepackage{bm}
\usepackage{amsmath}
\usepackage{aas_macros}
\newcommand{\comment}[1]{}

\def\om{$\Omega_m~$}

\def\mlim{M_{\rm lim}}

\begin{document}
\preprint{}

\title{Precision cosmology with a combination of\\ wide and deep Sunyaev-Zeldovich cluster surveys}

\author{Satej Khedekar$^1$, Subhabrata Majumdar$^1$ and Sudeep Das$^2$}
\affiliation{$^1$Tata Institute of Fundamental Research, Homi Bhabha Road, Colaba,
Mumbai - 400076, India.\\
$^2$Berkeley Center for Cosmological Physics, LBL and Department of Physics, University of California, Berkeley, California 94720, USA
}

\date{\today}

\begin{abstract}
We show the advantages of a wedding cake design for Sunyaev-Zel'dovich cluster surveys.
We show that by dividing up a cluster survey into a wide and a deep survey, one can essentially
 recover the cosmological information that would be diluted in a single survey of the same
 duration due to the uncertainties in our understanding of cluster physics. The parameter
 degeneracy directions of the deep and wide surveys are slightly different, and combining them
 breaks these degeneracies effectively. A variable depth survey with a few thousand clusters is
 as effective at constraining cosmological parameters as a single depth survey with a much larger
 cluster sample.
\end{abstract}

\maketitle

{\emph{Introduction.---}
The abundance and redshift distribution $\frac{dN}{dz}$ of clusters are  potentially important 
probes of cosmological parameters \citep{Holder01,
WangSteinhardt, Levine, Weller02, Hu}.
Recently, number counts from cluster surveys in optical (e.g RCS \citep{Gladders} and SDSS \citep{Rozo07})
 and microwave \citep{Vanderlinde} have been used to constrain cosmology. In the coming 
decade, dedicated
surveys in different wavebands (for example, ACT, DES, eROSITA, etc.) plan to use
clusters as probes of precision cosmology.  A correct cosmological interpretation of the 
observed $\frac{dN}{dz}$ depends on precise knowledge 
 of the limiting mass of the survey at a given redshift.  Unfortunately,  the mass of  a cluster
is not directly observable, but has to be obtained through  proxy
observables such  as X-ray surface brightness and temperature \citep{Ebeling00}, Sunyaev Zel’dovich (SZ) decrement 
\citep{Staniszewski, Hinks}, cluster richness
\citep{Postman, Koester} and lensing \citep{Wittman06, 2009arXiv0907.4232Z}. 
Uncertainties in the observable to mass conversion \comment{and redshift evolution}  create 
degeneracies
between cosmology and cluster physics and degrade cosmological constraints.
 There have been a number
of attempts in the literature to break such degeneracies, for example, through the
so called `self-calibration' techniques \citep{MM04, Hu, LimaHu04}. Other approaches 
include  an `unbiased' mass
follow-up of a subsample of the survey clusters \citep{MM03, MM04} or better 
theoretical modeling of clusters to predict the form of mass-observable scaling relation
\citep{YoungerHaiman, Reid06, CM09}. All of these are equivalent to putting
priors on the parameter space of the mass-observable relations. One can also try to 
optimize the cluster surveys so as to get the best possible survey yield \citep{Battye05}.

In this Letter, we show that by simply dividing the total observation
time between deep and wide surveys, parameter degeneracies can be effectively broken 
leading to tight cosmological constraints. This is without the need of any external
 information or a costly mass follow-up program. Throughout this paper, we assume a spatially
 flat cosmological model with a constant equation of state $w$ for dark energy with the
 fiducial cosmological parameters taken as $h$ = 0.717, $\Omega_{\rm{m}} = 0.258$,
 $w = -1$, $\Omega_{\rm{b}} = 0.044$, $n_s = 0.963$ and $\sigma_8 = 0.796$ consistent
 with the WMAP 5-year results. 

{\emph{Preliminaries.---}
The redshift distribution of detectable clusters in a survey over an area
$\Delta\Omega$ is,
\begin{equation}
\nonumber
\frac{dN}{dz}(z)=\Delta\Omega\frac{dV}{dzd\Omega}(z)\int_{0}^{\infty}f(M \rvert
 \mlim (z))\frac{dn(M,z)}{dM}dM
\label{eqn:dndz} ,
\end{equation}
where $\frac{dV}{dzd\Omega}$ is the comoving volume element and $\frac{dn}{dM}$
is the cluster mass function (taken from simulations by
\citet{Jenkins}). A mass limit $\mlim (z)$ is effectively imposed through the  
complementary error function $f(M \rvert \mlim (z))$ \citep{LimaHu05}, which is used
 to model a 20\% logarithmic scatter in the mass-proxy relation 
centered about $\mlim$. Figure \ref{fig:dndz} shows the plots of $\mlim$ versus $z$ for
 various array sensitivities of the ACT survey along with the corresponding redshift distribution of clusters.
 For an SZ survey, the flux-mass scaling relation is used to determine $\mlim (z)$ \citep{MM04},
${\rm{flux}_{\rm{sz}}(z,\nu)}d_{A}^{2}(z)={2.699\times10^{11}} g(\nu)A_{\rm SZ}M_{200}^{
\alpha}E^{2/3}(z)\left(1+z\right)^{\gamma}$,
 where $g(\nu)$ is the frequency dependence of the SZ distortion,
 $M_{200}$ is in units of
 $M_{\odot}$ and $d_{A}$ is the angular diameter distance in units of Mpc. 
 The Hubble expansion is  parameterized as $H(z)=H_0E(z)$.
The parameter $\gamma$ captures deviation in the redshift
 evolution of the cluster scaling relation from the self similar case, and we assume $\gamma 
\sim 0$.
 We choose the SZ scaling relations to be compatible with recent observations \citep
{Bonamente_scl, CM09}
with ${\rm log}(A_{\rm SZ})$ = -28.067 and $\alpha$ = 1.612. 

\begin{figure}[tb]
\begin{center}
\vspace*{-2 mm}
\includegraphics[width=3.4in]{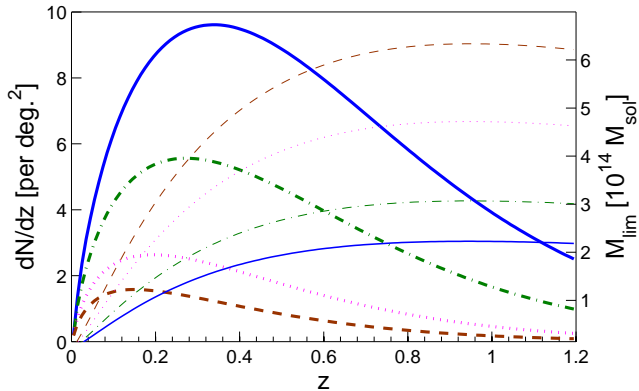}
\caption{The cluster redshift distribution $\frac{dN}{dz}$ {\it (thick lines)} and
$M_{\rm{lim}}$ {\it (thin lines)} vs redshift, for a SZ survey with array sensitivities of 
 3.54 {\it (solid)}, 10.61 {\it (dot-dashed)}, 21.21 {\it(dotted)} and 34.12 {\it(dashed)} $\rm{\mu K \sqrt{s}}$. \vspace*{-0.5 cm}}
\label{fig:dndz}
\end{center}
\end{figure}

\begin{figure}[b]
\begin{center}
\vspace*{-5 mm}
\includegraphics[width=3.5in]{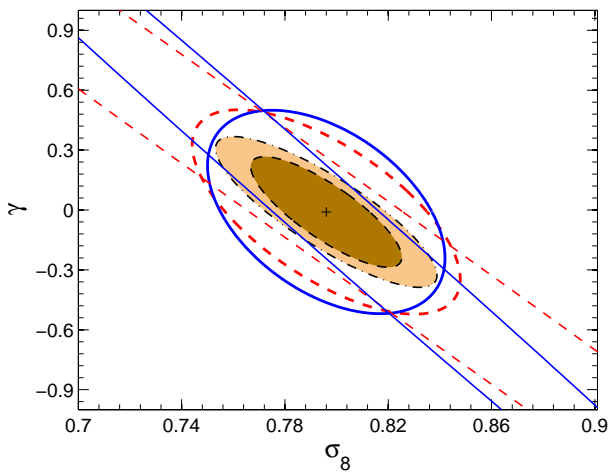}
\caption{Parameter degeneracy in the $\sigma_8$-$\gamma$ plane for a wedding cake survey
 with $f_{\rm{time}}$ = 0.75 and $r$ = 1/9. The {\it thick} ellipses indicate constraints when mass
follow-up is added, while the {\it thin} cigar-like ellipses are the constraints
for unknown cluster evolution. For these,
 the {\it dashed lines} are the constraints from the small area survey, while the
 {\it solid lines} are constraints from the large area survey.
The {\it light (dark)} filled ellipse shows the combined constraint from the two 
surveys with mass follow-up (unknown evolution). Note that, the intersection of the two cigars give the tighter constraints.}
\label{fig:degen}
\end{center}
\end{figure}

\citet{MM03, MM04} have shown that a substantial improvement in the cosmological 
parameter constraints can be obtained by adding a follow-up
of a fraction of the cluster sample to calibrate the mass-observable relation through detailed 
observations in X-Ray, SZ or galaxy spectroscopy. We construct a mock follow-up survey with $\sim 100$ clusters having 
`unbiased' mass measurements in the range $2-10 \times10^{14} h^{-1}$ $M_{\odot}$ 
between 0.3 $\leq$ z $\leq$ 1.1 with 20-40\% error on masses. We use both Fisher matrix and MCMC 
techniques, details on which  can be found in many references, e.g.
\cite{Tegmark, CosmoMC02, Holder01, MM04, Weller02,
Gladders}. Full MCMC runs are made to estimate the covariance matrix between cosmological parameters
and ($A_{SZ},\alpha,\gamma$) from follow-up. 
The full parameters space is spanned  by ($h, n_s, \Omega_{\rm{b}}, \Omega_m,
w,\sigma_8, A_{SZ},\alpha,\gamma$) and we 
use Gaussian priors with standard deviations of 0.027, 0.015 and 0.003 on
the parameters  $h$,  $n_s$ and $\Omega_{\rm{b}}$ respectively.

We assume the survey instrument to be a telescope with an angular resolution of  $\theta_
{\rm fwhm} $ with a  bolometric 
detector array of equivalent array-sensitivity (noise-equivalent temperature)  $\sigma_a$.
For a survey of area  $A$ and duration $t_{\rm obs}$, the time spent on each of the 
$N_{\rm{pix}} = A/\theta^{2}_{\rm{fwhm}}$ sky-pixels is  $t_{\rm{pix}} = t_{\rm obs}/N_{\rm{pix}}
$,  giving a noise per sky-pixel of $\Delta_T= \sigma_a/\sqrt{t_{\rm{pix}}}$.  In particular, we 
assume a survey of area $A=2000$ sq.~deg. at 150~GHz with $\theta_{\rm{fwhm}}=1^\prime$ 
and a duration of $10^7$  seconds.  We consider three cases for the array-sensitivity ---  
3.54, 10.61 and 21.21 $\rm{\mu K \sqrt s}$. In perspective, for its 2008 season, the ACT had an array sensitivity of $\sim 
34$  $\rm{\mu K \sqrt s}$ at 148 GHz. These array sensitivities would correspond to flux limits of 2.7, 8.3, 16.6
 and 26.7 $\rm{mJy}$ respectively for a single area of 2000 sq. deg. in the ACT survey. The SPT has similar configurations. Given a thermal noise $
\Delta_T$ per sky-pixel, the flux limit is computed as follows: we  assume that  the  
SZ decrement/increment has to be integrated over roughly 30 pixels to detect a cluster, so the 
$5\sigma$ flux limit becomes  $5\sqrt{30} \Delta_T$ converted from  $\rm{\mu K}$ into mJy through 
the derivative of the blackbody function \cite{WhiteMajumdar}. However, flux sensitivity may 
be  limited by the covariant noise from CMB as well as confusion arising from point sources;
we use a fixed (irrespective of how deep we go on a patch) limiting flux value of 2.5 mJy as an 
absolute lower cutoff.

\begin{figure}[b]
\begin{center}
\vspace*{-5 mm}
\includegraphics[width=3.4in]{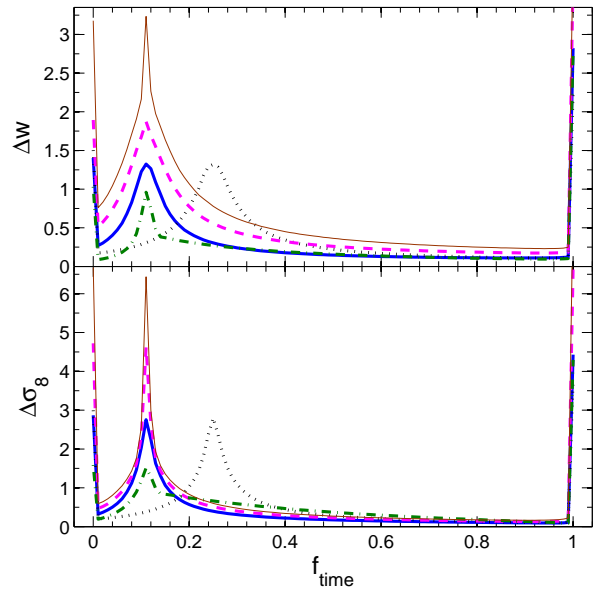}
\caption{Constraints on $w$ {\it (upper panel)}, and $\sigma_8$ {\it (lower panel)} from only number 
count observations for a two component survey design as a function $f_{\rm{time}}$.
 The curves are plotted for - $\sigma_a ( \rm{\mu K \sqrt s}) $ = 3.53, r =
1/9 {\it (dot-dashed)}; $\sigma_a$ = 10.61, r = 1/9 {\it (solid)}; $\sigma_a$= 21.21,
 r = 1/9 {\it (dashed)}, $\sigma_a$ = 10.61 \& r = 1/4 {\it (dotted)}, and $\sigma_a$ = 31.83 \& r = 1/9 {\it (thin)}.}
\label{fig:priors_w}
\end{center}
\end{figure}

\begin{figure}[b]
\begin{center}
\vspace*{-5 mm}
\includegraphics[width=3.5in]{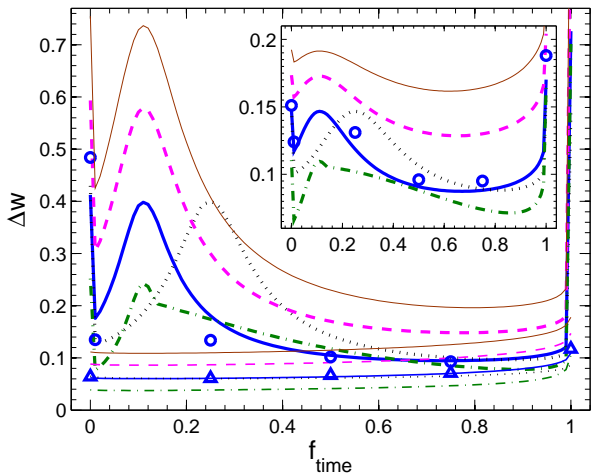}
\caption{Constraints on $w$ in a wedding cake survey. The main figure shows the 
constraints when ($A_{SZ}, \alpha$) are known. The {\it thick (thin)} curves correspond to an 
unknown (known) cluster evolution respectively. The inset displays constraints when ($A_{SZ}, \alpha, \gamma$) are free
 parameters but mass follow-up is added. The curves are plotted for $\sigma_a (\rm{\mu K \sqrt s})$ = 21.21, r = 1/9 
{(\it dashed)}; $\sigma_a$ = 10.61, r = 1/9 {\it (solid)}; $\sigma_a$ = 10.61, r = 1/4 {\it (dot-dashed)}, and $\sigma_a$ = 31.83 \& r = 1/9 {\it (thin)}.}
\label{fig:inset_w}
\end{center}
\end{figure}

{\emph{Results.---}

 We illustrate the strength of a wedding cake survey by 
breaking the fiducial 2000 deg$^2$  
survey area into two patches  keeping the total observing time
fixed. We use $f_{\rm{time}}$ to denote the
fraction of total time spent on the smaller area, and  $r$ to denote the ratio of the smaller to 
the larger survey area.  Varying $f_{\rm{time}}$ amounts to different integration time on each patch and 
hence different flux limits.\comment{  Fig.~\ref{fig:dndz} shows the
limiting mass $M_{\rm{lim}}$, along with the corresponding cluster
$\frac{dN}{dz}$ for various array sensitivities. Varying  $f_{\rm{time}}$}
Different flux limits lead to changes in the shape of $\frac{dN}{dz}$ (see Figure \ref{fig:dndz}), so 
that the two patches yield  different parameter degeneracy directions.  
When the constraints are combined, the parameter degeneracies are effectively broken, 
leading to tight constraints on cosmological parameters. This is evident in
 Fig.~\ref{fig:degen} which illustrates how the notorious cosmology-cluster 
physics degeneracy (such as $\sigma_8 - \gamma$)
 can be broken. A multiple depth survey is akin to having surveys with multiple limiting masses
or $\mlim$. This provides binning in both the observable mass proxy as well as redshift. {\it The binning in the mass
 is what helps in breaking the degeneracies, by extracting out more information from the survey} \citep{LimaHu05}.

{\emph{a) Wedding cake only.---}}
The marginalized constraints on $w$ and $\sigma_8$ from an SZ survey
 like the ACT for various detector noise levels as a function of
$f_{\rm{time}}$ are shown in Fig.~\ref{fig:priors_w}. It shows that having a single sky area to a 
fixed depth with no prior knowledge of cluster 
parameters gives no constraints on cosmology, in qualitative agreement with
previous work  \cite{MM04}. However, the constraints dramatically
improve as soon as the survey is broken up into wide and deep parts.
For our survey of total duration $10^7$ s and with cluster yield $N_{\rm{cl}} \sim
(3200,2700)$ for the wide and deep parts
(for $f_{\rm{time}} \sim 0.75$, $r=1/9$ and $\sigma_a= 3.54$ $\rm{\mu K \sqrt s}$) one can get 
 ($\Delta \Omega_m, \Delta w, \Delta \sigma_8$) =  (0.105, 0.115, 0.116) just
from $\frac{dN}{dz}$.  For a similar exercise, a single depth survey of 2000 deg$^2$ 
 with $N_{cl} \sim 7500$ would get  ($\Delta \Omega_m, \Delta w, \Delta
\sigma_8$) =  (0.772, 1.41, 3.31)!  {\it Thus,  moving to deep+wide survey acts like having an
additional mass follow-up (or `self-calibration') without any extra effort}. Note that the bump at 
$f_{\rm{time}}$ = 0.1 for 
$r=1/9$
corresponds to having  the same flux limits for the two
surveys such that the parameter degeneracies are the same and there is no
improvement in the constraints. For a higher area ratio, $r$ = 1/4,  the bump is
shifted to the right, as expected.

{\emph{b) Breaking degeneracies: Mass follow-up vs wedding cake survey.---}
Our lack of knowledge in the evolution in the mass-observable relations can lead to 
cosmological constraints getting weaker by factors of 4 or more  \cite{MM03}, which can be
restored  through either a mass follow-up program or using cluster bias
as a proxy for mass (i.e, self-calibration). However, the former requires a
costly follow-up effort and `unbiased' mass
estimators, while for the latter, one needs to be careful about scale dependent
bias. Here, we put forward a simple solution for this problem.  
Following \cite{MM04}, we parameterize the redshift evolution of mass - observable relation 
with a factor $(1+z)^{\gamma}$ in the scaling relation. To start, we assume that
$A_{SZ}$ and $\alpha$ are known perfectly. Fig.~\ref{fig:inset_w} shows the constraints on $w$ when $\gamma$ is known/unkown.
 We find that a two component survey improves the constraints on $w$ by a factor of 4 for
the case when cluster evolution is unknown. This is mainly by breaking the
 degeneracy existing between the parameters $\gamma$ and $\sigma_8$. As shown in 
Table \ref{tab:results}, the constraints on $w$ improve from 0.4 to 0.095 when we have a two component survey.
Note, that the results from MCMC (triangles and circles in Fig.~\ref{fig:inset_w} ) confirm the 
trends seen from the Fisher pipeline. The inset in Fig.~\ref{fig:inset_w} shows the constraints 
obtained on the parameter $w$ when both cluster structure and evolution is unknown but we have
some information on all the three cluster parameters from mass follow-up, the
currently preferred method for upcoming cluster surveys. The tightest constraints on $w$
are obtained when the smaller patch is observed
between 70\% and 95\% of the total time, depending of the array-sensitivity.
 Adding follow-up observations to cluster counts already breaks
the degeneracy in cosmology-cluster physics, but a wedding cake design can still improve
 $w$ constraints by a factor of 1.5 or more.
Again, we confirm the trend seen in Fisher matrix forecasts with MCMC simulations. A 
detailed comparison of Fisher-vs-MCMC for cluster
cosmology will be given in \citep{KM09}.

\begin{table}[t]\small
\caption{Comparison of $1\sigma$ parameter constraints from a cluster survey of
 array-sensitivity $10.61~\rm{\mu K\sqrt{s}}$ and duration $10^7$ seconds. The boldface numbers with
 similar brackets are for easy comparison of a {\it single component vs two component survey}.
\label{tab:params}  }
\hbox to \hsize{\hfil\begin{tabular}{cccc}
\hline
\hline
\multicolumn{3}{l}{\bf{Single area (2000 deg$^2$) survey}}\\
\hline
 & \multicolumn{1}{c}{unknown} & \multicolumn{1}{c}{unknown} & \multicolumn{1}{c}{with mass}  \\
Parameter \hspace{0.3cm} & \multicolumn{1}{c}{scaling} & \multicolumn{1}{c}{redshift evolution} & \multicolumn{1}{c}{follow-up}  \\
\hline
\hspace{0pt}$\Delta$\om		& 0.676 & 0.019 & 0.032 \\
\hspace{0pt}$\Delta w$ 	        & 1.343 & {\bf (0.400)} & {\bf [0.147]} \\
\hspace{0pt}$\Delta \sigma_8$ 	& 2.849 & 0.121 & 0.028 \\
\hline
\hline
\multicolumn{3}{l}{\bf{Wedding cake (1800+200 deg$^2$) survey}}\\
\hline
 & \multicolumn{1}{c}{unknown} & \multicolumn{1}{c}{unknown} & \multicolumn{1}{c}{with mass}  \\
Parameter \hspace{0.3cm} & \multicolumn{1}{c}{scaling} & \multicolumn{1}{c}{redshift evolution} & \multicolumn{1}{c}{follow-up}  \\
\hline
\hspace{0pt}$\Delta$\om		 & 0.105 & 0.009 &  0.030 \\
\hspace{0pt}$\Delta w$ 	         & {\bf [0.115]} & {\bf (0.095)} &  0.088  \\
\hspace{0pt}$\Delta \sigma_8$ 	 & 0.116 & 0.019 &  0.028 \\
\hline
\hline
\vspace{-1cm}
\end{tabular}\hfil}
\label{tab:results}
\end{table}

{\emph{Discussion.---}
We have shown that by breaking up a cluster survey 
into a wide and a deep parts, tight cosmological constraints can be obtained without 
the need for costly mass follow-up observations or self-calibration, which may be limited by our 
understanding of scale-dependent bias. For different scenarios
--- having only number counts, having a follow-up survey, or known scaling relations
with unknown redshift evolution, we always find an improvement in the
constraints on the dark energy equation of state $w$. Our finding are summarized in 
Table~\ref{tab:params}, which clearly shows that a two component survey is always superior 
to a single area survey. For an SZ survey like the ACT, just number count observations  with
 the two component design are capable of constraining cosmology very well ($\Delta \Omega_m$ = 0.105,
 $\Delta w$ = 0.115 and $\Delta \sigma_8$ = 0.116), when a single area approach would produce
 cosmologically uninteresting constraints. These results can be further improved by having follow-up
 observations of $\sim 100$ clusters. A wedding cake survey is also able to effectively break the
 degeneracy between non-standard evolution $\gamma$ and $\sigma_8$, and provide factors of several
 improvements in constraints over a single area survey (third column of Table~\ref{tab:params}).

With the advent of a large number of surveys in the near future, we
should be able to use clusters as independent probes of cosmology.
 Precision cluster cosmology is traditionally thought to 
require either a good understanding of cluster physics, or detailed mass 
 follow-up observations of a fraction of the detected clusters. We show that
 having a wedding cake survey strategy is an effective way to beat down
cosmological parameter degeneracies. Our work provides an important new insight
 into developing future SZ survey designs.

The authors would like to thank David Spergel for discussions. SM also thanks Anya Chaudhuri, Christoph Pfrommer
 and Joe Mohr for many discussions on scaling relations over the years.

\end{document}